\documentclass[english,superscriptaddress,floatfix,twocolumn,oneside,amsmath,amssymb,amsfonts,
aps,pra,floatfix]{revtex4-1}
\usepackage[english]{babel}
\usepackage[utf8x]{inputenc}
\usepackage{microtype}
\usepackage{graphicx}
\usepackage{siunitx}
\usepackage{xspace}
\usepackage[hidelinks]{hyperref}
\usepackage{bm}
\usepackage{bbold}
\usepackage{natbib}
\usepackage{csquotes}
\usepackage{physics}
\usepackage{xcolor}
\usepackage{qcircuit}

\usepackage[draft]{fixme}

\usepackage[capitalise]{cleveref}

\renewcommand{\vec}[1]{\ensuremath{\bm{#1}}\xspace}
\newcommand{\mat}[1]{\begin{bmatrix} #1 \end{bmatrix}}

\newcommand{\cnot}{\textsc{cnot}\xspace}
\newcommand{\swap}{\textsc{swap}\xspace}
\newcommand{\iswap}{$i$\swap}
\newcommand{\cz}{\textsc{cz}\xspace}

\newcommand{\python}{\textsc{Python}\xspace}
\newcommand{\pyscf}{\textsc{PySCF}\xspace}

\newcommand{\tfq}{\textsc{TensorFlow Quantum}\xspace}

\newcommand{\change}[1]{\textcolor{black}{#1}}

\newcommand{\affA}{Department of Physics and Astronomy, Aarhus University, DK-8000 Aarhus C, Denmark}
\newcommand{\affB}{Kvantify Aps, DK-2300 Copenhagen S, Denmark}
\newcommand{\affC}{Department of Chemistry, Aarhus University, DK-8000 Aarhus C, Denmark}

\date{\today}
\begin{document}
	
	\title{Parameterized Two-Qubit Gates for Enhanced Variational Quantum Eigensolver}
	
	\author{S. E. Rasmussen}
	\email{stig@phys.au.dk}
	\affiliation{\affA}
	\affiliation{\affC}
	\affiliation{\affB}
	\author{N. T. Zinner}
	\email{zinner@phys.au.dk}
	\affiliation{\affA}
	\affiliation{\affB}
	
	\begin{abstract}
		The variational quantum eigensolver is a prominent hybrid quantum-classical algorithm expected to impact near-term quantum devices. They are usually based on a circuit ansatz consisting of parameterized single-qubit gates and fixed two-qubit gates. We study the effect of parameterized two-qubit gates in the variational quantum eigensolver. We simulate a variational quantum eigensolver algorithm using fixed and parameterized two-qubit gates in the circuit ansatz and show that the parameterized versions outperform the fixed versions, \change{both when it comes to best energy and reducing outliers}, for a range of Hamiltonians with applications in quantum chemistry and materials science. 
	\end{abstract}
	
	\maketitle
	
	Hybrid quantum-classical algorithms (HQC) are likely to play an essential part in obtaining a quantum advantage on near-term NISQ devices \cite{Preskill2018,Endo2021,Bharti2022}. Among these algorithms, the variational quantum eigensolver (VQE) stands out as one of the most promising \cite{Peruzzo2014,McClean2016,Omalley2016,Kandala2017,Cao2019,Barkoutos2018,McCaskey2019,Gard2020,Cerezo2021,Motta2021}.
	The classical part of the VQE algorithm performs a variational optimization of the eigenvalues of a given Hamiltonian. The object of the quantum part is to supply trial states for the algorithm. A parameterized quantum circuit generates these trial states (PQC) called the circuit ansatz, depending on a set of control parameters. The classical part of the algorithm controls these parameters. There are numerous ways to design the circuit ansatz \cite{Kandala2017,McClean2016,Romero2018,Ostaszewski2021} which have led to many attempts trying to quantify the properties and capabilities of different PQC architectures \cite{Sim2019,Geller2018,Du2018,Benedetti2019,Hubregtsen2020}. However, in most cases, the circuit ansatz in a VQE algorithm combines parameterized single-qubit rotations and fixed two-qubit gates. 
	
	It is commonly believed, that may be beneficial to remove gates, adjust the amount of entangling gates, and remove some of them \cite{Kim2021}, and it has been shown that one can remove some single-qubit rotations \cite{Rasmussen2020c}. This paper follows this line of thought and investigates whether parameterized two-qubit gates improve the results of a VQE algorithm. \emph{A priori}, it might seem counter-intuitive that a possible reduction in entangling capability of a circuit should improve results. However, it is essential to remember that not all quantum systems are maximally entangled. Therefore, we hypothesize that a flexible amount of entanglement performs better than the maximal amount of entanglement for many systems in a VQE algorithm.

	The paper is organized as follows: \Cref{sec:vqe} gives a short introduction to the theoretical basis for the VQE algorithm. \Cref{sec:2qubitGates} discusses the two-qubit gates we will consider and how to parameterize them. \Cref{sec:results} shows the results of the VQE simulations for several different Hamiltonians. Finally, \cref{sec:conclusion} presents a summary and outlook of the findings.
	
	\section{Variational Quantum Eigensolver}\label{sec:vqe}
	
	The VQE algorithm is based on the variation principle, which gives an upper bound for the ground state energy, $E_0$, of some given Hamiltonian, $\hat H$. In general, it states that given any normalized state, $\ket\psi$, we have the following $E_0 \leq \mel{\psi}{\hat H}{\psi}$.
	In other words, the expectation value of the Hamiltonian will always be larger or equal to the actual ground state energy, no matter what trial state we choose. 
	We are interested in approximating the eigenstates of the Hamiltonian using trial states in a way such that $\ket*{\psi(\vec \theta)}$ depends on a set of parameters $\vec \theta$. In that case, we can minimize the expectation value of the Hamiltonian in order to approximate the ground state energy.
	
	The idea behind the VQE is that a quantum circuit might be better at choosing trial states than a classical computer. Therefore, computing the trial states is done on a quantum processor while calculating expectation values, and minimizing the energy is done on a classical computer. To create a parameterized quantum state we choose a circuit ansatz, represented by the operator $\hat U(\vec \theta)$, such that $\ket*{\psi(\vec \theta)} = \hat U(\vec \theta) \ket{0}$.
	
	In order to find multiple eigenstates of the Hamiltonian, We employ subspace-search Variational Quantum Eigensolver (SSVQE) \cite{Nakanishi2019}.
	Using just a single optimization procedure, this can find up to the $k$th excited state. The idea behind the procedure is that instead of using a single input state as $\ket 0 = \ket{0\dots 0}$ we choose a set of orthogonal states, $\{\ket \varphi_j\}_{j=0}^k$, meaning that $\braket*{\varphi_i}{\varphi_j} = \delta_{ij}$, where $\delta_{ij}$ is the Kronecker delta and then minimize a cost function consisting of the sum of matrix elements given by these states. In other words, we minimize the following function
	\begin{equation}
	F(\vec \theta) =  \sum^{k}_{j=0} \eta_i\mel*{\varphi_j}{\hat U(\vec \theta)^\dagger \hat H \hat U(\vec \theta) }{\varphi_j},
	\end{equation}
	where $\eta_i$ is some weight between 0 and 1 satisfying $\eta_i > \eta_j$ when $i < j$, which chooses which state becomes the excited states. When optimized this cost function maps the $\ket*{\varphi_j}$ to the $j$th excited state for each $j \in \{ 0, 1, \dots , k\}$. While this procedure only requires one optimization step, the overall time required for optimization may increase.
	
	In an actual VQE implementation, one can create a set of orthogonal states by applying a Pauli-X operation to a set of different qubits all initiated in the state $\ket 0$. For example applying a Pauli-X gate to the first qubit yields $\textsc{x}_1 \!  \ket{0\dots 00} = \ket{0 \dots 01}$, which is orthogonal to the original state.

	\section{Two-qubit Entangling Gates}\label{sec:2qubitGates}
	
\begin{figure}
		\centering
		\begin{equation*} \Qcircuit @C=1em @R=0.5em @!R  {
			& \qw & \ctrl{1} & \qw & \ctrl{1} & \qw & \qw & \qw  \\
			& \gate{R_z(\frac{\pi}{2})} & \targ & \gate{R_y(-\frac{\Theta}{2})} & \targ & \gate{R_y(\frac{\Theta}{2})} & \gate{R_z(-\frac{\pi}{2})} & \qw  \\
		}
		\end{equation*}
		\caption{\change{Standard gate decomposition of the parameterized $\cnot(\Theta)$.}}
		\label{fig:paramCNOT}
	\end{figure}
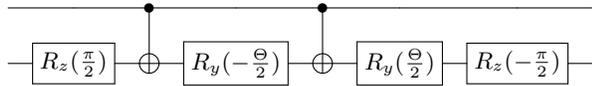

	There are different approaches to choosing the best circuit ansatz \cite{Kandala2017,McClean2016,Romero2018,Ostaszewski2021}; however, typical for all of them; it is usually built from single-qubit rotations and fixed two-qubit entangling gates. The logic behind this is that the single-qubit rotations provide a way to parameterize the circuit, while the two-qubit gate entangles the circuit. Nevertheless, there is no reason why these two-qubit gates should not be parameterized as well. The two-qubit gate is usually chosen to have maximal entangling power; however, it is not clear that trial states benefit from a forced maximum entangling power. Single-qubit rotations are cheap to implement experimentally, and we do not propose to substitute these rotations but rather to supplement with parameterized two-qubit gates. Most experimental implementations of two-qubit gates require some tuning, and parameterizing this tuning instead of keeping it fixed creates a parameterized two-qubit gate at potentially very little experimental cost.
	In superconducting circuits this tuning of the two-qubit gates is usually performed using microwave driving similarly to how single-qubit rotations are driven \cite{Lucero2008,Chow2009,Motzoi2009,Lucero2010,Johnson2015,McKay2017,Rasmussen2020a,Rigetti2010,Chow2011,Rasmussen2021} or using flux tuning \cite{McKay2016,Yan2018,Kounalakis2018,Sung2020,Li2020,Rasmussen2020b}. Even for the case where parameterized gates are difficult to perform directly on a given quantum platform one can easily transpile the parameterized two-qubit gates into fixed two-qubit gates and single qubit gates, e.g., the parameterized $\cnot(\Theta)$ in \cref{eq:cnotParam} is equivalent to the circuit in \cref{fig:paramCNOT}.

	We, therefore, investigate whether using parameterized two-qubit gates will improve VQE results. In particular, we consider three different parameterized entangling gates: the $\cnot(\Theta)$, $i $$\swap(\Theta)$, and the $\cz(\Theta)$ gates, which are defined in the following way 
	\begin{subequations}\label{eq:paramEntGates}
		\begin{align}
		\cnot(\Theta) &= \mat{1 & 0 & 0 & 0\\ 0 & 1 & 0 & 0 \\ 0 & 0 & \cos(\Theta/2) & -ie^{i\Theta/2}\sin (\Theta/2) \\ 0 & 0 & -ie^{i \Theta/2}\sin (\Theta/2) & \cos (\Theta/2) }, \label{eq:cnotParam} \\
		i\swap(\Theta) &= \mat{1 & 0 & 0 & 0 \\ 0 & \cos (\Theta/2) & -i\sin(\Theta/2) & 0 \\ 0 & -i\sin(\Theta/2) & \cos(\Theta/2) & 0 \\ 0 & 0 & 0 & 1}, \\
		\cz(\Theta) &= \mat{1 & 0 & 0 & 0\\ 0 & 1 & 0 & 0 \\ 0 & 0 & 1 & 0 \\ 0 & 0 & 0 & e^{i\Theta}},
		\end{align}
	\end{subequations}
	where we denote the angle parametrizing the coupling by $\Theta$. We consider these three gates since these are the parameterized versions of the most commonly used entangling gates. Further, the $i$$\swap(\Theta)$ and $\cz(\Theta)$ gates are easily implemented in superconducting circuits using capacitive coupling \cite{Krantz2019,Yan2018,Sung2020,Li2020,McKay2016,Rasmussen2020b,Chen2021,Rasmussen2021}. Note that for $\Theta = 0$, the gates are the identity, while for $\Theta = \pi$, they yield their standard fixed version. The entangling power \cite{Williams2011} increases monotonously between these two extremes as a sigmoid curve.
	
	Another example of a tunable two-qubit gate that can be natively implemented in superconducting circuits is the cross-resonance gate \cite{Rigetti2010,Chow2011}. This gate has been used in VQE experiments in its fixed version \cite{Kandala2017}, but it is straightforward to use it in its tunable version.
	While we have not used this gate in our VQE simulations, it should behave similarly to the gates used here as it can be transformed to the universal \cnot gate using single-qubit rotations.
	
	\section{Results}\label{sec:results}
	
	\begin{figure}
		\centering
		\includegraphics{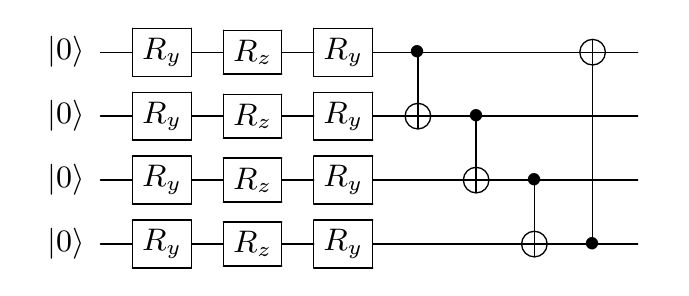}
		\caption{Example of one layer of the circuit ansatz, $\hat U_l(\vec \theta_l)$, for four qubits. First, an Euler rotation is performed on each qubit, followed by a nearest-neighbor coupling using the given entangling gate. Here it is shown using fixed \cnot gates. \change{For each qubit three parameters are related to the Euler rotation, while a single parameter is related to the two-qubit gate, if it is parameterized.}}
		\label{fig:ParamGatesCircuit}	
	\end{figure}
	
	To see the effect of parameterized two-qubit gates in variational quantum algorithms we simulate an SSVQE implemented using the \python package \tfq \cite{Broughton2021}. \change{Throughout the paper we will consider a noiseless model. A noisy model is beyond the scope of this paper, however, we do note that a noisy VQE potentially could outperform a noiseless VQE \cite{Mueller2022}. We optimize the parameterized quantum circuits using the \textsc{Adam} optimizer \cite{Kingma2014} with a learning rate of 0.1. This stochastic gradient-based algorithm is one of the most commonly used in the field of machine learning due to its computational efficiency. The use of other optimizers would possibly change the results of a given optimization problem, since we are considering a rather large parameter landscape, however, sampling several times will make up for this.}

	We choose a layered circuit ansatz such that $\hat U(\vec \theta) = \hat U_L(\vec \theta_L) \cdots \hat U_2(\vec \theta_2)\hat U_1(\vec \theta_1)$, where each layer, $\hat U_l(\vec \theta_l)$, consists of an Euler rotation on each qubit followed by nearest-neighbor couplings of all qubits using two-qubit gates. \change{Such a circuit ansatz could be realized with a linear qubit layout, such as employed, e.g., by Rigetti's Aspen devices or by IBM Q. In general nearest-neighbor couplings is a rather sparse coupling, which makes it realizable on many qubit layouts.}
	
	This means that we have $3N$ single-qubit rotations in each layer, where $N$ is the number of qubits. We have an additional $N$ parameters for the parameterized two-qubit gates, yielding $4N$ parameters for each layer, compared to $3N$ parameters for the fixed two-qubit gates.
	It might seem like an unfair comparison with more parameters for the circuits with parameterized two-qubit gates compared to the fixed circuit, given them an advantage. However, it should not be a problem as the circuits are already saturated with parameters \cite{Rasmussen2020c}. Nonetheless we even out the playing field following the same vein and randomly remove a single-qubit gate for each qubit such that the all circuits have $3N$ parameters per layer.
	
	An example of a layer can be seen in \cref{fig:ParamGatesCircuit}, where the layer is shown with \cnot gates. However, we simulate for both the fixed and parameterized versions of the gates in \cref{eq:paramEntGates}.

	We simulate the SSVQE applied to several different Hamiltonians typically used to benchmark technical developments with VQEs. In the following subsection, we present a sample of the results and refer to the Supplementary Material for additional results on the performance.
	
	\subsection{Molecular Hamiltonians}
	
	\begin{figure}
		\centering
		\includegraphics[width=.85\columnwidth]{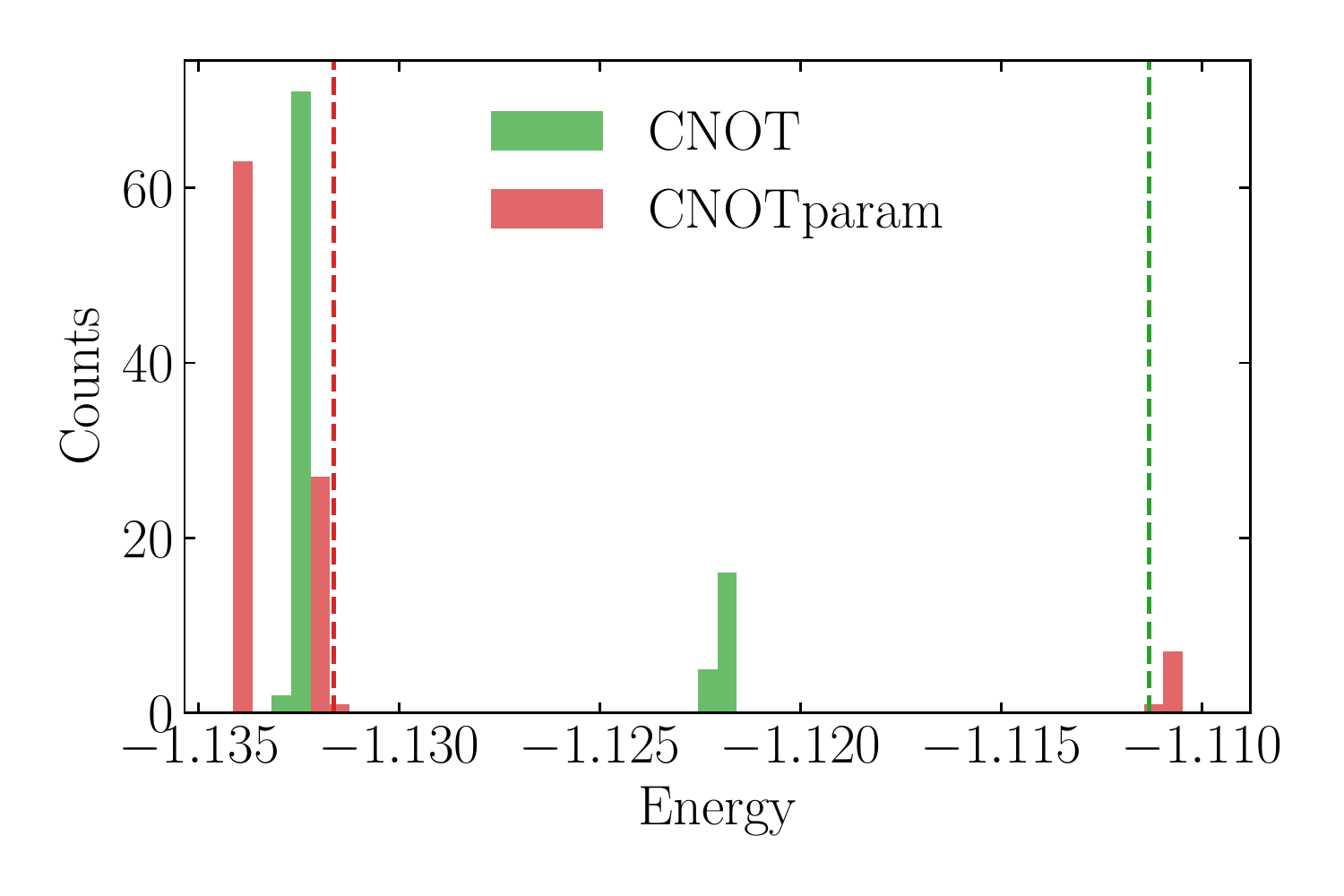}
		\caption{\change{Histogram of the results from 100 calculations of the H$_2$ molecule with 2 layers and fixed or parameterized \cnot gates. The dashed line indicate the average values of the calculations. For the fixed \cnot gate there are a few outliers around $-0.8$ and $-0.6$, which are not seen in the plot, which is also why its average values is further to the right. Generally we observe that the energies fall in bands, which is due to the banded energy structure of molecules. Other calculations yields similar results.}}
		\label{fig:histo}
	\end{figure}

	\begin{figure*}[ht]
		\centering
		\includegraphics[width=.9\textwidth]{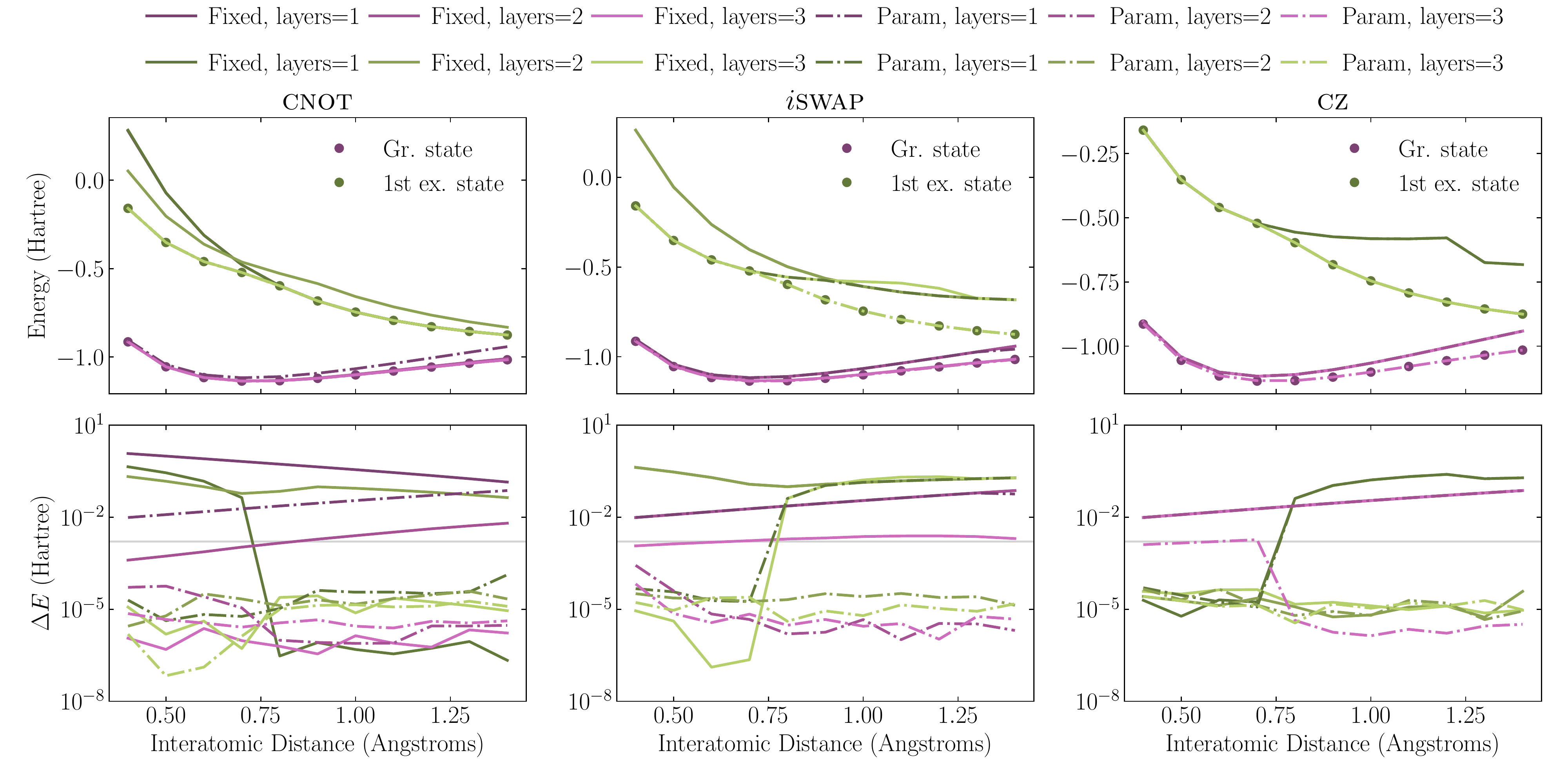}
		\caption{Potential energy surfaces for H$_2$ calculated using SSVQE for the two lowest states. The title of each column indicates which entangling gate is used in the calculation, either fixed (solid line) or parameterized (dash-dotted lines). In the top row, we show the calculated energy, and in the bottom row, we show the energy difference, $\Delta E$, between the classical and VQE calculations on a log-scale. Lighter colors indicate that more layers are used in the simulation. The gray line indicates chemical accuracy at $0.0016$ Hartree. The results shown here are the best out of 100 samples.}
		\label{fig:H2}
	\end{figure*}
	
	Small molecular Hamiltonians are often chosen as benchmarks for VQEs as they present an accessible real-world problem \cite{Kandala2017,Peruzzo2014,Omalley2016,Cao2019,McCaskey2019,McArdle2020,Bauer2020,Arute2020,Elfving2021}, and is the precursor of larger molecular Hamiltonians which we would ultimately like to solve using VQEs. Therefore, we consider H$_2$, LiH, and BeH$_2$ since their active space can be effectively represented by 4, 4, and 6 qubits. In order to apply the VQE to molecular problems, the Hamiltonians must be encoded onto qubits. We do this using the Jordan-Wigner \cite{Jordan1928} transformation. Following such a transformation, a Hamiltonian for, e.g., four qubits takes the form
	\begin{equation}
	\hat H = \sum_{\substack{\alpha,\beta, \gamma,\delta \\ \in x, y, z, I}} h_{\alpha\beta\gamma\delta} \sigma_1^\alpha \sigma_2^\beta \sigma_3^\gamma \sigma_4^\delta,
	\end{equation}
	where $\sigma_i^{x,y,z}$ are the Pauli operators and $\sigma_i^I$ is the identity on qubit $i=1,2,3,4$. The coefficients $h_{\alpha\beta\gamma\delta}$ are given by one- and two-body integrals of the molecule and do generally depend on the molecular basis. We find the fermionic Hamiltonian using the Python-based Simulations of Chemistry Framework (\pyscf) \cite{pyscf2018}. The matrix elements are calculated in the STO-3G basis, and we map the Hamiltonian to a qubit Hamiltonian using OpenFermion \cite{OpenFermion2017}. This procedure can, in principle, be applied to all molecules for all geometries; however, we restrict ourselves to small molecules to avoid diverging computation time and issues of classical optimization. \change{One of the main issues of the classical optimization part of HQC algorithms is barren plateaus. These can potentially be fixed using quantum constitutional neural networks (QCNN) \cite{Pesah2021}. However, QCNNs are beyond the scope of this paper.}
	
	We simulate the molecules for varying interatomic distances to obtain a potential energy surface for the two lowest-lying states for each molecule. We perform each calculation 100 times, \change{the distribution of results for one calculation of the ground state can be seen in \cref{fig:histo}. The spread in the results happen since each calculation is started with random initial conditions and we are using a stochastic optimization algorithm. Generally we observe that the non-parameterized versions tend to have outliers far above the best results, while the parameterized versions do not have these outliers. We also observe that the parameterized versions return the lowest best energy.} 
	
	\change{However, the comparison in \cref{fig:histo} is not fair since the parameterized version have more parameters. We therefore compare results for a different number of layers, in order to see whether parameterized two-qubit gates needs less layers than fixed versions. Since we are doing variational calculations, we compare the calculation resulting in the lowest energy.} We compare both the VQE energy and the difference, $\Delta E$, between the VQE energy and the energy found by exact diagonalization of the qubit Hamiltonian.

	In \cref{fig:H2} we show the results for the two lowest states of H$_2$, plotting results for up to three layers for both fixed and parameterized entangling gates. 
	Starting with the \cnot gate, we see that the parameterized gates outperform the fixed gates for one and two layers, while the performance is somewhat identical for three layers. We note that we achieve chemical accuracy \cite{Kandala2017} using only two layers for the parameterized gate, compared to three for the fixed gates. Note the sharp drop in $\Delta E $ for the fixed gate for one layer around $\SI{0.75}{\angstrom}$; this happens since here we have a crossing of states, where the H$_2^+$ state crosses the triplet state of H$_2$. The SSVQE cannot distinguish between these states and find the lowest. However, the one-layer fixed gate only finds the second-lowest to begin with, and since the starting values of the VQE come from the previous calculation, it overperforms when this crossing occurs. \change{If one wishes to find a state of a given multiplicity or charge, and thus fix the problem of the curve crossing, one must employ a different type of VQE, such as the constrained VQE \cite{Ryabinkin2019}.}
	
	The story is the same for the \iswap gate, and we even see that the two-layer parameterized circuits perform better than fixed three-layer circuits. We are not capable of achieving chemical accuracy with just fixed gates. This time we see a sharp increase in $\Delta E$ around $\SI{0.75}{\angstrom}$ for two of the excited states. Again it is due to the crossing; the algorithm finds the lowest and continues with this state.
	Finally, for the \cz gate, we find that it performs significantly worse than the two other gates, both for the fixed and parameterized versions. We only achieve chemical accuracy for the ground state for three layers of the parameterized gate. However, the \cz gate works surprisingly well for the fixed and parameterized versions for the first excited state, except for one layer, where the crossing is again the source of trouble. We believe that the \cz gate is better at approximating the excited state than the ground state because the \cz gate mimics the structure of the first excited state better. 
	We show similar results for LiH and BeH$_2$ in the Supplementary Material. For LiH we see that chemical accuracy is achieved for just a single layer of parameterized versions of \cnot and \iswap, while the fixed version need more layers. For the \cz gate we observe no difference between the fixed and parameterized version. For BeH$_4$ we see that the parameterized version of the \cnot gate outperform the fixed version for the ground state energy when using two layers, but for the excited state the results are identical for both cases. For the \iswap gate we see improved performance for the parameterized version when using two layers. However we observe slightly decreased performance for three layers. For the \cz gate we observe no significant difference between the parameterized and fixed versions.

	\subsection{Heisenberg Model}
	
	\begin{figure*}[ht]
		\centering
		\includegraphics[width=.9\textwidth]{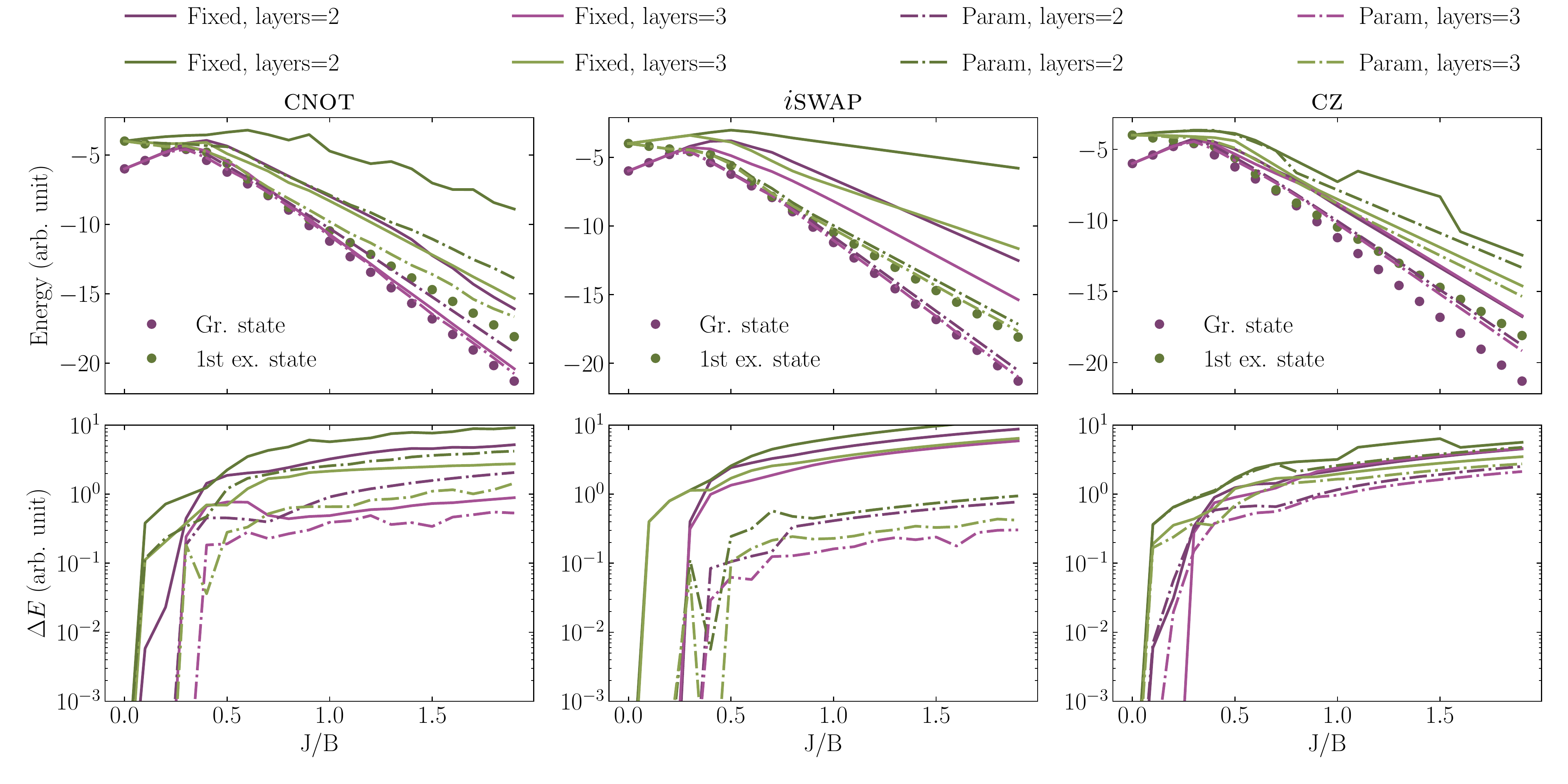}
		\caption{Potential energy surfaces for the Heisenberg model in \cref{eq:Heisenberg} with six qubits calculated using SSVQE for the two lowest states. The title of each column indicates which entangling gate is used in the calculation, either fixed (solid line) or parameterized (dash-dotted lines). In the top row, we show the calculated energy, and in the bottom row, we show the energy difference, $\Delta E$, between the classical and VQE calculations on a log-scale. Lighter colors indicate that more layers are used in the simulation. The results shown here are the best out of 100 samples.}
		\label{fig:Heisenberg}
	\end{figure*}
	
	We also consider a problem of quantum magnetism, namely a one-dimensional Heisenberg model with periodic boundary conditions in the presence of an external magnetic field, i.e.,
	\begin{equation}\label{eq:Heisenberg}
	\hat H = B \sum_{i=1}^N \sigma_i^z  + J \sum_{\langle ij \rangle} \left( \sigma^x_i \sigma^x_{j} + \sigma^y_i \sigma^y_{j} + \sigma^z_i \sigma^z_{j} \right).
	\end{equation}
	As with the molecular Hamiltonians, we consider a range of configurations; more specifically, we consider $J/B \in [0, 2]$ varied in steps of 0.1. It is an interesting case for parameterized entangling gates since for $J=0$, the ground state is entirely separable, and for increasing $J$, the ground state becomes increasingly entangled.

	In \cref{fig:Heisenberg} we consider the Heisenberg model with $N=6$ qubits.
	We note that the VQE model is very good at determining the energy when $J=0$, which makes sense since the Hamiltonian is diagonal in the computational basis in this case. The ground state is also nicely determined until the crossing around $J/B = 0.3$, at which point the fixed gates start performing worse than the parameterized gates. In other words, the parameterized two-qubit gates are better at mimicking this change in entanglement than the fixed versions.
	Especially for the \iswap gate, we observe that the parameterized version outperforms the fixed version. The first excited state is more difficult for the VQE, and it generally performs poorly, although the parameterized version of the \iswap performs best of all the gates.
	Here we also see that a two-layer PQC with parameterized gates is better than the three-layer fixed version, i.e., it is not just the increased number of parameters in the model that yields a better result.
	Altogether, the Heisenberg model results align with the results for the molecular Hamiltonian. In the Supplementary Material, we present results for the Heisenberg model with four, eight, and ten qubits. These results are consist with the ones present here for six qubits: The parameterized versions generally outperform the fixed versions and the parameterized \iswap gate is the best performing gate.
	
	\subsection{Transverse-field Ising Model}
	
	\begin{figure*}[ht]
		\centering
		\includegraphics[width=.9\textwidth]{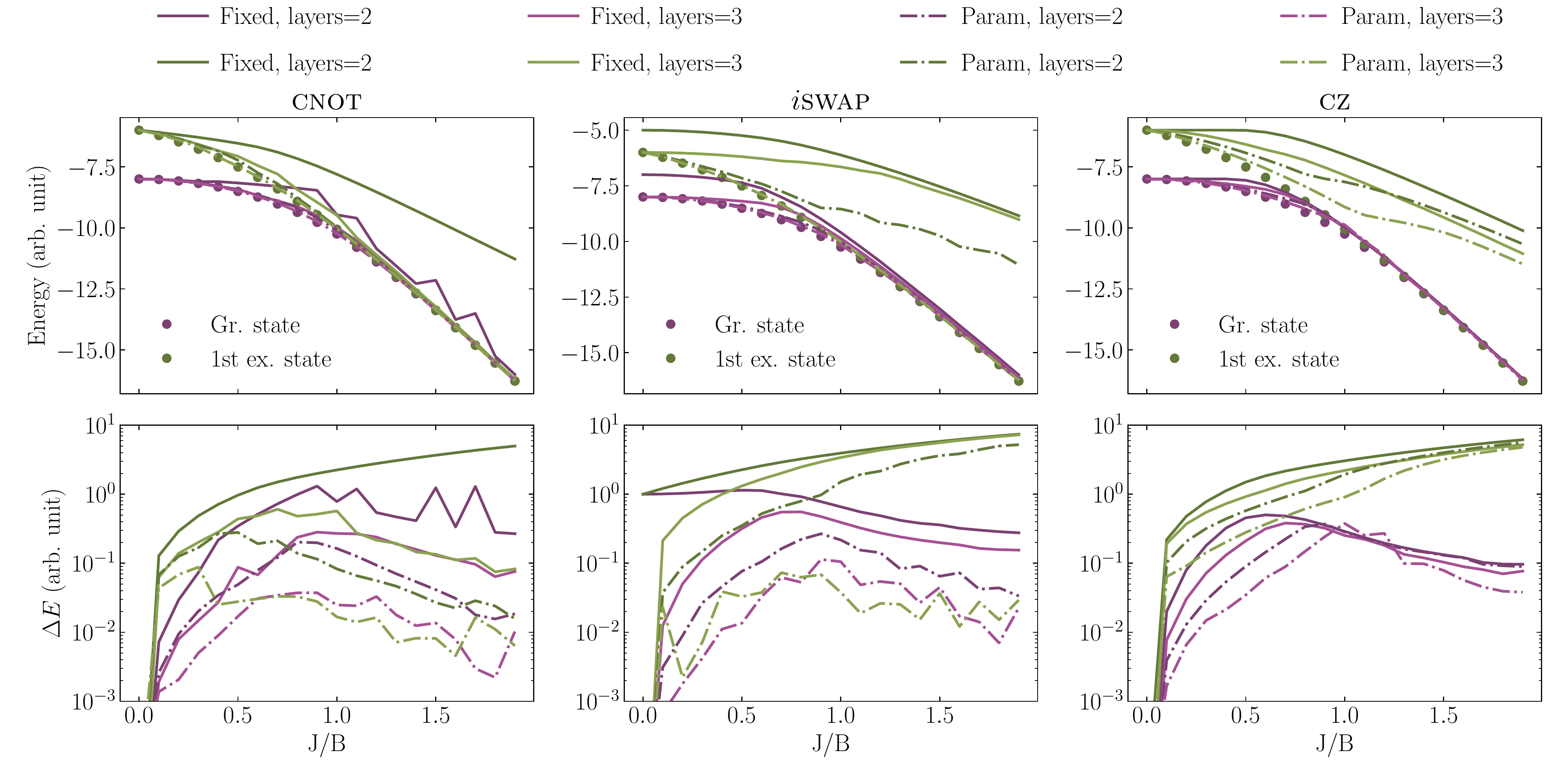}
		\caption{Potential energy surface for the transverse-field Ising model in \cref{eq:TransFieldIsing} with eight qubits calculated using SSVQE for the two lowest states. The title of each column indicates which entangling gate is used in the calculation, either fixed (solid line) or parameterized (dash-dotted lines). In the top row, we show the calculated energy, and in the bottom row, we show the energy difference, $\Delta E$, between the classical and VQE calculations on a log-scale. Lighter colors indicate that more layers are used in the simulation. The results shown here are the best out of 100 samples.}
		\label{fig:Ising}
	\end{figure*}
	
	Finally, we consider the transverse-field Ising model, which features interaction along the $z$ axis and an external magnetic field perpendicular to the $z$ axis. It has the Hamiltonian
	\begin{equation}\label{eq:TransFieldIsing}
	\hat H = B\sum_{i=1}^N \sigma_i^x  + J \sum_{\langle ij \rangle} \sigma^z_i \sigma^z_{j} .
	\end{equation}
	In \cref{fig:Ising} we show results for simulations of the transverse-field Ising model for $N=8$ qubits, as with the Heisenberg model, the ground state is completely separable for $J=0$ but becomes increasingly entangled for increasing $J$. The conclusion is the same: The parameterized versions of the two-qubit gates perform as good or better than the fixed versions. For the \cnot and \iswap, we see that the parameterized versions perform significantly better for fewer layers than their fixed versions.
	In the Supplementary Material, we present results for the Transverse-field Ising model with four, six, and ten qubits. These results are identical to the ones presented here: The parameterized versions generally performs better than the fixed versions, and for the \cnot and \iswap gates fewer layers even yield a better performance when using the parameterized version of a given gate.
	
	\section{Conclusion and outlook}\label{sec:conclusion}
	
	We have investigated how parameterized two-qubit gates affect the performance of the variational quantum eigensolver compared to their fixed counterparts usually used in VQE algorithms. We considered several different Hamiltonians; Molecular, Heisenberg, and transverse-field Ising models, all analyzed using the SSVQE framework, to show that our results are not model and state specific. 
	We considered a layered circuit ansatz consisting of Euler rotations on each qubit followed by nearest-neighbor interactions between each qubit. A point for further investigation could be to test parameterized two-qubit gates in a different circuit ansatz, e.g., all-to-all coupled circuits.
	
	In general, the parameterized version of two-qubit gates performs better than the fixed version, and sometimes it is even more beneficial to use a parameterized two-qubit gate than increasing the number of layers. In other words, better performance is achieved with fewer parameters. 
	It should not, \emph{a priori}, be surprising that parameterizing previously fixed gates in a parameterized quantum circuit should increase the part of the Hilbert space available to the PQC, thus improving VQE results. Nevertheless, parameterized two-qubit gates seems to be rarely used in hybrid quantum-classical algorithms. This is regrettable since they are often just as easy to implement experimentally as their fixed versions. Put in another way; it is a free, untapped quantum resource, which could improve near-term quantum devices.
	
	Finally, we note that our results are in good agreement with Ref. \cite{Yordanov2020} who shows that unitary coupled cluster ansatz' can be significantly reduced using controlled rotation gates, i.e., the parameterized \cnot gate and varieties of this. Reference \cite{Yordanov2020} also mention that the ansatz could be improved using multi-qubit.controlled rotations gates, something which could be effectively implemented in superconducting circuits \cite{Rasmussen2020a}.

	\begin{acknowledgements}
		The authors thank M. Majland and N. J. S. Loft for discussion on different aspects of the work.
		This work is supported by the Danish Council for Independent Research. Some of the numerical results presented here were obtained at the Centre for Scientific Computing, Aarhus \url{http://phys.au.dk/forskning/cscaa/}.
	\end{acknowledgements}

	\clearpage
%	\bibliography{ParamGates}
	
\end{document}